\def\H{{\sqrt{\Delta}}}

\def\diff#1#2{\frac{\partial #1}{\partial #2}}
\def\Hh{{\mathcal H}}

\documentclass[12pt,oneside]{amsart}

\usepackage{amsmath}               % AmSLaTeX
\usepackage{amsthm}                % aggiunge nuovi ambienti tipo teorema
\usepackage{times}
\usepackage{graphicx}
\theoremstyle{plain}\newtheorem{theorem}{Theorem}[section]
\theoremstyle{plain}\newtheorem{lemma}[theorem]{Lemma}
\theoremstyle{plain}\newtheorem{proposition}[theorem]{Proposition}
\theoremstyle{plain}
\theoremstyle{definition}\newtheorem{definition}[theorem]{Definition}
\theoremstyle{remark}

\numberwithin{equation}{section}

\title[the Cosmic Censor to the court]%
{New solutions of Einstein equations in spherical symmetry: the
Cosmic Censor to the court}

\author[R.\ Giamb\`o ,\ F.\ Giannoni]{Roberto Giamb\`o, Fabio Giannoni}
\address{Dipartimento di Matematica e Informatica,
Universit\`a di Camerino, Italy}
\email{roberto.giambo@unicam.it, fabio.giannoni@unicam.it}
\author[G.\ Magli]{Giulio Magli}
\address{Dipartimento di Matematica, Politecnico di
Milano, Italy}
\email{magli@mate.polimi.it}
\author[P.\ Piccione]{Paolo Piccione}
\address{Dipartimento di Matematica e Informatica,
Universit\`a di Camerino, Italy \hfill\break\indent
(on leave from Departamento de Matem\'atica,\hfill\break\indent
Universidade de S\~ao Paulo, Brazil)}
\email{paolo.piccione@unicam.it, (piccione@ime.usp.br)}

\begin{document}

\begin{abstract}
A new class of solutions of the Einstein field equations in
spherical symmetry is found. The new solutions are mathematically
described as the metrics admitting separation of variables in
area-radius coordinates. Physically, they describe the
gravitational collapse of a class of anisotropic elastic
materials. Standard requirements of physical acceptability are
satisfied, in particular, existence of an equation of state in
closed form, weak energy condition, and existence of a regular
Cauchy surface at which the collapse begins. The matter properties
are generic in the sense that both the radial and the tangential
stresses are non vanishing, and the kinematical properties are
generic as well, since shear, expansion, and acceleration are also
non-vanishing. As a test-bed for cosmic censorship, the nature of
the future singularity forming at the center is analyzed as an
existence problem for o.d.e. at a singular point using techniques
based on comparison theorems, and the spectrum of endstates -
blackholes or naked singularities - is found in full generality.
Consequences of these results on the Cosmic Censorship conjecture
are discussed.
\end{abstract}

\maketitle

\section{Introduction}\label{sec:intro}
The final state of gravitational collapse is an important open
issue of classical gravity. It is, in fact, commonly believed that
a collapsing star that it is unable to radiate away - via e.g.
supernova explosion - a sufficient amount of mass to fall below
the neutron star limit, will certainly and inevitably form a black
hole, so that the singularity corresponding to diverging values of
energy and stresses will be safely hidden - at least to faraway
observers - by an event horizon. However, this is nothing more
than a conjecture - what Roger Penrose first called a "Cosmic
Censorship" conjecture \cite{pe} - and has never been proved. On
the contrary, in the last twenty years of research, many analytic
examples of (spherically symmetric) naked singularities satisfying
the principles of physical reasonableness have been discovered.

Spherically symmetric naked singularities can be divided into two
groups: those occurring in scalar fields models \cite{Ch94,Ch99}
and those occurring in astrophysical sources modeled with
continuous media, which are of exclusive interest here (see
\cite{review} for a recent review). The first (shell focusing)
examples of naked singularities where discovered numerically by
Eardley and Smarr \cite{e2} and, analytically, by Christodoulou
\cite{e3}, in the gravitational collapse of dust clouds. Since
then, the dust models have been developed in full details: today
we know the complete spectrum of endstates of the gravitational
collapse of spherical dust with arbitrary initial data \cite{sj}.
This spectrum can be described as follows: given the initial
density and velocity of the dust, a integer $n$ can be introduced,
in such a way that if $n$ equals one or two, the singularity is
naked, if $n\geq 4$ the singularity is covered, while if $n=3$ the
system undergoes a transition form naked singularities to
blackholes in dependence of the value of a certain parameter. If
the collapse is marginally bound, the integer $n$ is the order of
the first non vanishing derivative of the energy density at the
center.

The dust models can, of course, be strongly criticized from the
physical point of view. Since from one side very few results are
known on perfect (i.e. isotropic) fluids and, on the other side,
stresses have to be expected to be anisotropic in strongly
collapsed objects, one of us initiated some years ago a program
whose objective was to understand what happens if the
gravitational collapse occurs in presence of only one
non-vanishing stress, the tangential one \cite{m1,m2}. The program
can be said to have been completed: we know the complete spectrum
of the gravitational collapse with tangential stresses in
dependence of the data both for the clusters \cite{ha,jmc} and for
the continuous media models \cite{gjm}.

The results of this program are somewhat puzzling. In fact, what
one should expect on physical grounds is that the equation of
state plays a relevant role in deciding the final state of the
collapse. Well, it is only {\it apparently} so. In fact, also in
the tangential stress case a parameter $n$ can be defined, in such
a way that the endstates of collapse depend on $n$ exactly in the
same way as in dust.

In the present paper we present a complete, new model of
gravitational collapse which includes both radial and tangential
stresses. This is done deriving a class of anisotropic solutions
which is in itself new, and contains the dust and the tangential
stress metrics as special cases. As far as we are aware, this is
the first time that the spectrum of the endstates of a solution
satisfying all the requirements of physical reasonability and
exhibiting both radial and tangential stresses is found.

As will be discussed in the concluding section, in view of the
results of the present paper, the case for a cosmic censor  - at
least in spherical symmetry - becomes very weak.

\section{Einstein Field Equations in area--radius coordinates}

Consider a spherically symmetric collapsing object. The general
line element in comoving coordinates can be written as
\begin{equation}\label{eq:ds-tr}
\text ds^2=-e^{2\nu}\text dt^2 +(1/\eta)\,\text dr^2 +R^2 (\text
d\theta^2 +\sin^2\theta\, \text d\phi^2)
\end{equation}
(where $\nu, \eta$ and $R$ are functions of $r$ and $t$). We shall
use a dot and a prime to denote derivatives with respect to $t$
and $r$ respectively.

In the present paper we consider as admissible sources of the
gravitational field only those matter models which admit a
well-defined thermodynamical description in terms of the standard
relativistic mechanics of continua. We recall that the physical
properties of an elastic material in isothermal conditions can be
described in terms of one state function $w$, which depends on
three parameters (the so called strain parameters) and on the
coordinates. In the comoving description, the strain tensor of the
material can be expressed in a "purely gravitational" way and, as
a consequence, the state function depends only on the space-space
part of the metric(see e.g. \cite{kmq} and references therein). In
addition, in spherical symmetry, the state function cannot depend
on angles, and therefore the equation of state of a general
spherically symmetric material can be given as a function of $r$
and of two strain parameters \cite{m1}:
\begin{equation}\label{eq:intnrg}
w=w(r,R,\eta) \ .
\end{equation}
It is useful also to introduce the matter density
\begin{equation}\label{eq:rho}
\rho =\frac{\sqrt\eta}{4\pi ER^2}
\end{equation}
(where $E=E(r)$ is an arbitrary positive function) so that the
internal energy density is given by $\epsilon=\rho w$. If $w$
depends only on $\rho$, one recovers the case of the barotropic
perfect fluid; in general, however, the stresses are anisotropic
and are given by the following stress-strain relations:
\begin{align}\label{eq:pr-ph}
&p_r=2\rho\eta\frac{\partial w}{\partial\eta},
&p_t=-\frac12\rho R\frac{\partial w}{\partial R}.
\end{align}

Comoving coordinates $r,t$ are extremely useful in dealing with
gravitational collapse because of the transparent physical meaning
of the comoving time. We shall, however, make systematic use here
also of another system of coordinates, the so-called area-radius
coordinates, which were first introduced by Ori \cite{Ori} to
study charged dust, and then successfully applied to other models
of gravitational collapse (see e.g. \cite{ha,m2}). These
coordinates prove extremely useful for technical purposes, as will
be clear below. In these coordinates the comoving time is replaced
by $R$. The velocity field of the material
$v^\mu=e^{-\nu}\delta^\mu_t$ transforms to $v^\lambda=e^{-\nu}\dot
R \delta^\lambda_R$ and therefore the transformed metric, although
non diagonal, is still comoving. It can be written as
\begin{equation}\label{eq:ds}
\text ds^2=-A\text dr^2 -2B\text dR\text dr -C \text dR^2 +R^2
(\text d\theta^2 + \sin^2\theta\,\text d\varphi^2)\ ,
\end{equation}
where $A$, $B$ and $C$ are functions of $r$ and $R$. Obviously
\begin{equation}\label{eq:C}
C=\frac 1{v_\lambda v^\lambda}:=\frac 1{u^2}
\end{equation}
where we have denoted $u=|\dot R e^{-\nu}|$. Formula
\eqref{eq:intnrg} implies that the internal energy $w$ now depends
on two coordinates ($r$ and $R$) and on only one field variable
$\eta$. It is convenient to introduce the quantity
\begin{equation}\label{eq:Delta}
\Delta := B^2-AC=\frac{1}{\eta u^2}>0\ ,
\end{equation}
(so that $\sqrt\eta=1/(u\H)$ and $B=-\sqrt{\Delta+A/u^2}$), and to
use $A,u$ and $\H$ as the fundamental field variables. A
convenient set of Einstein equations for these is $G^r_r=8\pi
T^r_r$, $G_r^R=8\pi T_r^R$ and $G_R^r=8\pi T_R^r$ \cite{m2}.
Denoting partial derivatives with a comma, the first of these
equations reads
\begin{equation}\label{eq:efe1}
\frac 1{R^2}\left[ 1-\frac A\Delta -R\left(\frac A\Delta\right)_{,R}
\right] =-8\pi\,p_r\ ,
\end{equation}
so that $A$ decouples from $u$ and $\Delta$:
\begin{equation}\label{eq:A-expr}
A=\Delta\left[1-\frac{2F(r)}R -\frac{2\Lambda}R\right],
\end{equation}
where
\begin{equation}\label{eq:Lambda}
\Lambda:=-\int_{R_0(r)}^R 4\pi \sigma^2 p_r (r,\sigma,\eta
(r,\sigma))\,\text d\sigma.
\end{equation}
In the above formulae, the arbitrary positive function $R_0(r)$
describes the values of $R$ on the initial data and will be
conveniently taken to be
\begin{equation}\label{eq:R0}
R_0(r)=r.
\end{equation}
To find the physical meaning of the other arbitrary function $F$,
we introduce the so called {\it Misner-Sharp mass} $\Psi(r,R)$. It
is defined in such a way that the equation $R=2\Psi$ spans the
boundary of the {\it trapped region}, i.e. the region in which
outgoing null rays reconverge. This boundary is the {\it apparent
horizon} of the spacetime, and it can be shown that (see e.g.
\cite{jbook}):
\begin{equation}\label{eq:Psi1}
\Psi (r,R)=(R/2)\left(1- g^{\mu\nu}\partial_\mu R\,\partial_\nu
R\right).
\end{equation}
Transforming to radius-area coordinates and using
\eqref{eq:A-expr}, one easily gets
\begin{equation}\label{eq:Psi}
\Psi =(R/2)\left(1-A/\Delta\right)=F(r)+\Lambda
\end{equation}
and therefore the function $F$ is the value of the Misner-Sharp
mass on the data. Since $\Lambda$ vanishes if the radial stress is
zero, when $p_r=0$ the Misner-Sharp mass is conserved during the
evolution (i.e. independent on $R$). This is the main technical
reason which makes the spacetimes with vanishing radial stresses
simpler with respect to general collapse models.

Using \eqref{eq:A-expr} it can be shown that the two remaining
field equations can be written as follows:
\begin{equation}\label{eq:efe2a}
E\,\Psi_{,r}=\left(w+2\eta \diff w\eta
\right)\sqrt{u^2+1-\frac{2\Psi}R}
\end{equation}
and
\begin{equation}\label{eq:efe3a}
(\sqrt{\Delta})_{,R}=-\frac 1{\sqrt{u^2+1-\frac{2\Psi}R
}}\left(\frac 1u\right)_{,r}\ .
\end{equation}

\section{The general solution admitting separation of variables}

\subsection{The solution}\label{subsec:sol}

Let us consider the system of two coupled PDE's for $\Delta$ and
$u$ \eqref{eq:efe2a}--\eqref{eq:efe3a}. If $\Lambda$ does not
depend on the field variables, equation \eqref{eq:efe2a} becomes
algebraic and the system decouples. Functional properties of
$\Lambda$ are related to the choice of $w$ and in fact, $\Lambda$
will be independent of $\eta$ if $p_r$ does. This obviously
happens if $p_r$ is zero (leading to the already well known cases
of dust and of  vanishing radial stresses, for which $w$ does not
depend on $\eta$) but also if $p_r$ is some function of $r$ and
$R$ only. This in turn occurs if $w$ is of the form
\begin{equation}\label{eq:w}
w(r,R,\eta)=h(r,R)+\frac 1{\sqrt\eta}\, l(r,R),
\end{equation}
the case $p_r=0$ corresponding to $l=0$. If $l$ is non zero the
radial pressure is  non-vanishing
\begin{equation}\label{eq:pr2}
p_r=-\frac 1{4\pi ER^2}\,l(r,R)
\end{equation}
so that
\begin{equation}\label{eq:Lambda-now}
\Lambda=\frac 1{E}\int_{r}^R l(r,\sigma)\,\text d\sigma.
\end{equation}

Using \eqref{eq:w} and \eqref{eq:Psi}, equation \eqref{eq:efe2a}
gives
\[
E\,\Psi_{,r}=h\sqrt{u^2+1-\frac{2\Psi}R},
\]
which  allows us to compute the quantity $u^2$:
\begin{equation}\label{eq:efe2b}
u^2(r,R)=-1+\frac{2\Psi}R+\left(\frac {E\Psi_{,r}}h\right)^2.
\end{equation}
As a consequence, \eqref{eq:efe3a} becomes a quadrature which
allows the calculation of $\sqrt\Delta$:
\begin{equation}\label{eq:sqrtDelta}
\sqrt\Delta=-\int\frac
1{Y(r,\sigma)}\diff{\Hh}{r}(r,\sigma)\,\text d\sigma,
\end{equation}
where we have defined
\begin{equation}\label{eq:Hh}
\Hh(r,R)=\sqrt{\frac{R}{2\Psi(r,R)+R(Y^2(r,R)-1)}},
\end{equation}
\begin{equation}\label{eq:Y}
Y(r,R)=\frac{E(r)\Psi_{,r}(r,R)}{h(r,R)}.
\end{equation}
It is convenient to eliminate the indefinite integral in
\eqref{eq:sqrtDelta}. For this aim, it is useful the following
equation
\begin{equation}\label{eq:idY}
Y(r,R)=R'(r,t(r,R))\sqrt\eta(r,t(r,R)),
\end{equation}
that can be easily found from the relations between the metrics
coefficients in the comoving and the area--radius coordinates.
Using \eqref{eq:Delta}, \eqref{eq:idY}
and the fact that $R'=1$ for $R=r$, we get
$\sqrt\Delta(r,r)=\frac{\Hh(r,r)}{Y(r,r)}$, which fixes the
integration constant, to obtain
\begin{equation}\label{eq:Delta-expr}
\sqrt\Delta(r,R)=\int_R^r\frac{1}{Y(r,\sigma)}\diff\Hh
r(r,\sigma)\,\text d\sigma+\frac{\Hh(r,r)}{Y(r,r)}.
\end{equation}

\subsection{Physical requirements}\label{subsec:phys}

In this section we will discuss the conditions that have to be
imposed on the equation of state and on the data, in order for the
solutions to be physically meaningful.

From now on, we consider only spacetimes whose matter source
satisfies the equation of state \eqref{eq:w}. For such spacetimes,
using \eqref{eq:Psi} and \eqref{eq:Lambda-now}, the constitutive
function $w$ may be written as
\begin{equation}\label{eq:w-now}
w(r,R,\eta)=E(r)\left(\frac{\Psi_{,r}(r,R)}{Y(r,R)}+\frac
{\Psi_{,R}(r,R)}{\sqrt\eta}\right),
\end{equation}
therefore the energy density $\epsilon=\rho w$ and the stresses
\eqref{eq:pr-ph} are given by:
\begin{align}
&\epsilon=\frac{1}{4\pi
R^2}\left(\frac{\sqrt\eta}{Y}\Psi_{,r}+\Psi_{,R}\right),\label{eq:eps-now}
\\
&p_r=-\frac{\Psi_{,R}}{4\pi R^2},\label{eq:pr-now}\\
&p_t=-\frac{\sqrt\eta}{8\pi
R}\left(\frac{\Psi_{,rR}}{Y}-\frac{\Psi_{,r}}{Y^2}\diff
YR+\frac{\Psi_{,RR}}{\sqrt\eta}\right), \label{eq:pt-now}
\end{align}

From now on we choose as independent arbitrary functions $\Psi$
and $Y$. In the present paper, we assume $C^k$ regularity of the
data: the equation of state and the arbitrary functions are
assumed to be Taylor-expandable up to the required order. Due to
eq. \eqref{eq:eps-now}, \eqref{eq:pr-now} and \eqref{eq:pt-now},
it will then be possible to translate all the conditions to which
the metrics have to satisfy in order to be physically meaningful
(like e.g. weak energy condition or regularity of the center up to
singularity formation, see below), in terms of conditions on the
functions $\Psi$ and $Y$ or on their derivatives.

We first impose the weak energy condition (w.e.c.). In spherical
symmetry, this condition is equivalent to $\epsilon\ge 0$,
$\epsilon+p_r\ge 0$, $\epsilon+p_t\ge 0$. These inequalities lead
to
\begin{equation}\label{eq:wec}
{\Psi_{,r}}\ge 0,\quad\Psi_{,R}\ge 0,\quad \Psi_{,r} \geq \frac R2
Y \left(\frac{\Psi_{,r}}{Y}\right)_{,R} ,\quad \Psi_{,R} \geq
\frac R2 \Psi_{,RR} .
\end{equation}
For an explicit solution satisfying weak energy condition
we refer to the example in \ref{subsec:aniso} below.

We also impose regularity of the metric at the center (`local
flatness'). In comoving coordinates $r,\,t$ this amounts to
require
\begin{equation}\label{eq:reg}
R(0,t)=0,\qquad e^{\lambda(0,t)}=R'(0,t).
\end{equation}
In addition, the stress tensor must be isotropic at $r=0$, that is
\begin{equation}\label{eq:preg}
p_r(0,t)=p_t(0,t),
\end{equation}
for any regular $t={\rm const}$ hypersurface. Finally, we require
the existence of a regular Cauchy surface ($t=0$, say) carrying
the initial data for the fields. These requirements are
fundamental, since they assure that the singularities eventually
forming will be a genuine outcome of the dynamics.

We have already chosen $R=r$ at $t=0$; using \eqref{eq:eps-now},
\eqref{eq:idY} and the fact that $R'(r,0)=1$ we find the
expression for the initial energy density $\epsilon_0(r)$:
\begin{equation}\label{eq:eps0}
\epsilon_0(r)=\frac{\Psi_{,r}(r,r)+\Psi_{,R}(r,r)}{4\pi r^2}.
\end{equation}
Once the regularity conditions are satisfied, the data will be
regular if this function is regular. For physical reasonability we
also require the initial density to be decreasing outwards, that
is $\epsilon_0'(r)\le 0$ for $r>0$. It is easy to check, that the
above stated conditions are equivalent, in terms of the functions
$\Psi$ and $Y$, to the following:
\begin{align}
&Y(0,0)=1,\label{eq:cond2}\\
&\Psi(0,0)=\Psi_{,r}(0,0)=\Psi_{,R}(0,0)=\Psi_{,rr}(0,0)=
\Psi_{,rR}(0,0)=\Psi_{,RR}(0,0)=0,\label{eq:cond1}\\
&\Psi_{,rr}(r,r)+2\Psi_{,rR}(r,r)+\Psi_{,RR}(r,r)-\frac 2r
\left(\Psi_{,r}(r,r)+\Psi_{,R}(r,r)\right)<0\label{eq:cond3}.
\end{align}

\begin{definition}\label{def:1}
We say that a spacetime  is a physically valid area-radius
separable spacetime (ARS) if
\begin{enumerate}
\item the equation of state of the matter is
\eqref{eq:w-now}, where $\Psi(r,R)$ and $Y(r,R)$ are arbitrary
positive functions;
\item the weak energy condition, the regularity condition, and the
condition of decreasing initial density hold, i.e. $\Psi(r,R)$ and
$Y(r,R)$ satisfy \eqref{eq:wec} and \eqref{eq:cond2}--\eqref{eq:cond3}.
\end{enumerate}
\end{definition}

\subsection{Kinematics}
Spherically symmetric non-static solutions can be invariantly
classified in terms of their kinematical properties (see e.g.
\cite{Kra,K}). Most of the known solutions have vanishing shear,
or expansion, or acceleration. For the solutions studied here all
such parameters are generally non-vanishing.

The acceleration in comoving coordinates $r,t$ is given by $a_\mu=
\nu'\delta_\mu^r$, therefore, it can be uniquely characterized by
the scalar $A:=\sqrt{a_\mu a^\mu}$. Using \eqref{eq:idY} together
with the relation
\begin{equation}\label{eq:nuprime}
Y\nu'=R'Y_{,R}
\end{equation}
(which easily follows from the
the field equation in comoving coordinates $\dot R'=\dot R
\nu'+R'\dot\lambda$) we get
\begin{equation}\label{eq:acc}
A=Y_{,R}.
\end{equation}
The expansion $\Theta$ is given by
\begin{equation}\label{eq:exp}
\Theta= \pm u\left[\frac{(u\sqrt\Delta)_{,R}}{u\sqrt\Delta}+\frac
2R\right].
\end{equation}
Finally, the shear tensor $\sigma_\nu^\mu $ can be uniquely
characterized by the scalar $\sigma:=(2/3) \sqrt{ \sigma_\nu^\mu
\sigma^\nu_\mu }$ given by
\begin{equation}\label{eq:shear}
\sigma=\pm \frac u3\left[\frac 1R-
 \frac{(u\sqrt\Delta)_{,R}}{u\sqrt\Delta}\right].
\end{equation}

\subsection{Special classes}

Special cases of the solutions discussed above are:

\begin{enumerate}

\item The dust (Tolman-Bondi) spacetimes. The energy
density equals the matter density, and this implies $\Psi_{,R}=0$
and $Y=E\Psi_{,r}$.

\item The general solution with vanishing radial stresses. Vanishing of
$p_r$
implies $\Psi_{,R}=0$ (see \eqref{eq:pr-now}), while $Y$  depends
also on $R$ (if $Y_{,R}=0$ one recovers dust). The properties of
these solutions have been widely discussed in \cite{m1}.

\item
An interesting new subclass is obtained by imposing the vanishing
of the acceleration. This subclass corresponds to a choice of the
function $Y$ depending on $r$ only. It is well known that
acceleration-free perfect fluid models can describe only very
special collapsing objects (the pressure must be a function of the
comoving time only) while these anisotropic solutions exhibit
several interesting features, like e.g. a complete spectrum of
endstates \cite{progr}.
\end{enumerate}

\subsection{An explicit example: "anisotropysations'' of
Tolman--Bondi--de Sitter spa\-ce\-time}\label{subsec:aniso}

The investigation of various explicit examples of new solutions
within the class presented here, as well as their physical
applications, will be presented in a forthcoming paper
\cite{progr}. We restrict ourself here only to sketch a simple
acceleration free (i.e. $Y=Y(r)$) case, essentially with the aim
of showing explicitly that the new sector of the solutions (w.r.
to dust and tangential stress case) is non empty.

A simple way to produce explicit examples of new collapsing
solutions within the equation of state considered here is to
choose the mass function as the sum of a function of $r$ only
(governing the dust limit) and a function of $R$ only (governing
the anisotropic stresses). In order to satisfy the requirements of
Definition \ref{def:1}, easy calculations show that the function
$\Psi$ must have the form
\begin{equation}\label{eq:TBdeS-ani}
\Psi(r,R)=\int_0^r\phi(s)s^2\,\text ds+\int_0^R\chi(\sigma)\sigma^2\,\text d\sigma,
\end{equation}
where $\chi$ and $\phi$ are positive and non increasing functions.
In the particular case in which $\chi$ is a constant, the
contribution of $\chi$ to the energy-momentum tensor is formally
the same as that of a cosmological constant, and therefore the
spacetime coincides with the so called Tolman--Bondi--de Sitter
(TBdS), describing the collapse of spherical dust. This is
actually the unique model of gravitational collapse in presence of
a lambda term which is known so far. In recent years, the
description of gravitational collapse in such spacetimes attracted
a renewed interest both from the astrophysical point of view,
since recent observations of high-redshift type Ia supernovae
suggest a non-vanishing value of lambda, and from the theoretical
point of view, after the proposal of the so-called Ads-Cft
correspondence in string theory. In such a context, the solutions
\eqref{eq:TBdeS-ani} can be used to investigate the effects of
stresses by adding higher order terms to the function $\chi$. For
instance, choosing
$$
\Psi(r,R)=\alpha r^3 + \frac{\Lambda R^3}{1+R/R_*}
$$
where $R_*$ is a positive constant, one obtains a model which is
TBdS homogeneous and isotropic near the center but becomes
anisotropic and inhomogeneous when $R$ is of the order of $R_*$
(we stress, however, that this is only one possible application of
this special sub-class of solutions).

\section{Conditions for singularity formation}\label{sec:sing}

\subsection{Shell crossing and shell-focusing singularities}\label{subsec:formation}

Due to eqs.\eqref{eq:Delta} and \eqref{eq:eps-now}, the energy
density becomes singular if, during the evolution, $R$ or
$u\sqrt\Delta$ vanish. The latter case corresponds to $R'(r,t)=0$
in comoving coordinates and is called a {\sl shell--crossing}
singularity, since it is generated by shells of matter
intersecting each other. Shell-crossing singularities correspond
to weak (in Tipler sense) divergences of the invariants of the
Riemann tensor. As a consequence, these singularities are usually
considered as "not interesting", although no proof of
extensibility is as yet present in the literature. In any case, we
shall concentrate here on the the shell-focusing $R=0$
singularity, for which no kind of extension is possible.
Therefore, we work out the conditions ensuring that crossing
singularities do not occur.

Since we are considering a collapsing scenario, $R<r$ during
evolution. Therefore, due to eq. \eqref{eq:Delta-expr}, it is
sufficient to require strict positivity of the function
$\sqrt\Delta(r,0)$ together with non-increasing behavior of
$\sqrt\Delta(r,R)$ w.r. to $R$. We thus have the following:
\begin{proposition}\label{prop:nosc}
In a ARS spacetime the formation of shell--crossing singularities
does not occur if
\begin{equation}\label{eq:Hrpos}
\Psi_{,r}(r,R)+Y(r,R)\,Y_{,r}(r,R)\,R\ge 0 \quad\text{for\ \ }
r\ge 0,\,R\in[0,r],
\end{equation}
\begin{equation}\label{eq:nosc}
\int_0^r\frac{1}{Y(r,\sigma)}\diff\Hh r(r,\sigma)\,\text
d\sigma+\frac{\Hh(r,r)}{Y(r,r)}> 0\qquad\text{for\ \ }r>0.
\end{equation}
\end{proposition}

The above condition, although being only sufficient, allows for a
wide class of new metrics. It becomes also necessary in the
particular case of dust \cite{lake}.

Spherically symmetric matter models do not, of course, unavoidably
form singularities if stresses are present. One can, in fact,
construct models of oscillating or bouncing spheres. In comoving
coordinates $(r,t)$, the locus of the zeroes of $R(r,t)$ - if any
- defines implicitly a singularity curve $t_s(r)$ via
$R(r,t_s(r))=0$. The quantity $t_s(r)$ represents the comoving
time at which the shell labelled $r$ becomes singular, and
therefore the central singularity forms if
\begin{equation}\label{eq:t0}
\lim_{r\to 0^+} t_s(r)=t_*  ,
\end{equation}
is finite ($t_*$ is positive due to the regularity of the data at
$t=0$ ). To study the behavior of this limit in dependence of the
choice of the data, we use $\dot R=-e^\nu u$. Integrating along a
flow line we get
\begin{equation}\label{eq:t}
t=\int_R^r e^{-\nu(r,\sigma)} \Hh(r,\sigma) \text d\sigma,
\end{equation}
where the initial condition \eqref{eq:R0} has been used. Up to
time repa\-ra\-mete\-ri\-za\-tions we can assume $\nu(0,t)=0$.
Using this fact, together with \eqref{eq:nuprime},
it can be seen that $\nu$ is uniformly bounded
and $\lim_{r\to 0^+}\nu(r,\sigma)=0$ uniformly for $\sigma\in[0,r]$.
Therefore it is necessary and sufficient to require
that, for $r$ near 0, $\Hh (r,\sigma )$ is integrable with respect to $\sigma$
in $[0,r]$ and
\begin{equation}\label{eq:integral2}
0<\lim_{r\to 0^+}\int_0^r \Hh (r,\sigma ) d\sigma <\infty .
\end{equation}
For this aim, let us consider the following Taylor expansion
centered at the point $(0,0)$, where we take Definition \ref{def:1} into account:
\begin{equation}\label{eq:H}
2\Psi(r,R)+R(Y^2(r,R)-1)=2Y_{,r}(0,0)r\,R+2Y_{,R}(0,0)R^2+\sum_{k\ge 3}\sum_{i+j=k}h_{ij} r^i
R^j.
\end{equation}
Changing variable $\sigma=r\tau$ and recalling the definition of
$\Hh$ (eq. \eqref{eq:Hh}) the integral in \eqref{eq:integral2}
becomes
\[
\int_0^1\frac{\sqrt\tau\,\sqrt r}{
\sqrt{2Y_{,r}(0,0)\tau +2Y_{,R}(0,0)\tau^2+ r\,
\sum_{i+j=3}h_{ij}\tau^j+r\,\varphi(r)}}\,\text
d\tau,
\]
with $\varphi(0)=0$. Convergence of this integral to a finite
non-zero value requires vanishing of the zero order terms and at
least one non vanishing first order term in the denominator (i.e.
$(h_{30},h_{21},h_{12})\not=(0,0,0)$). We shall, however, consider
only the sub-case in which $h_{30}$ (which we will call $\alpha$
hereafter) is non-vanishing, since a vanishing $\alpha$ would
correspond to a bad-behaved "dust limit" of the solutions, i.e.
when the source of the anisotropic stress becomes very weak. For
the same reason, $\alpha$ cannot be negative otherwise the weak
energy condition would be violated in the same limit. Thus:
\begin{proposition}\label{prop:sf}
In an ARS spacetime shell focusing singularities form if
\begin{equation}\label{eq:Yder}
Y_{,r}(0,0)=Y_{,R}(0,0)=0, \  \alpha>0 .
\end{equation}
\end{proposition}

From the above discussion, we now define the subclass of ARS
spacetimes that we are dealing with hereafter.
\begin{definition}\label{def:2}
An ARS spacetime is called collapsing if:
\begin{enumerate}
\item Shell-crossing singularities do not form (i.e. $\Psi$ and $Y$
satisfy to \eqref{eq:Hrpos}, \eqref{eq:nosc});
\item Shell-focusing singularities form in a finite amount of
comoving time (i.e. $\Psi$ and $Y$ satisfy to \eqref{eq:Yder}).
\end{enumerate}
\end{definition}

\subsection{The apparent horizon and the nature of the possible
singularities}

A key role in the study of the nature of a singularity is played
by the apparent horizon (see for instance \cite{jbook}). The
apparent horizon in comoving coordinates is the curve $t_h(r)$
defined by $R(r,t_h(r))=2\Psi(r,R(r,t_h(r)))$. In area-radius
coordinates, one has correspondingly a curve $R_h(r)$ and, since
$\Psi_{,R}(0,0)=0$, implicit function theorem ensures that
$R_h(r)$ is well defined in a right neighborhood of $r=0$ and such
that $R_h(r)<r$. This fact, along with the requirement
$\Psi_{,R}\ge 0$ coming from the w.e.c. \eqref{eq:wec} ensures
$\Psi(r,R_h(r))<\Psi(r,r)$. Now using \eqref{eq:t} it is
\begin{equation}
\label{eq:ts-th} 0\le t_s(r)-t_h(r)= \int_0^{2\Psi(r,R_h)}e^{-\nu}
\Hh(r, \sigma) d\sigma\le
\int_0^{\Psi(r,r)}e^{-\nu}\Hh(r,\sigma)d\sigma
\end{equation}
Changing variable  $\sigma=r\tau$ as before and using the fact
that $\nu$ is bounded and $\Psi(0,0)=0$,
we find that this integral converges to 0 as
$r$ tends to 0, ensuring the following
\begin{proposition}\label{prop:centre}
In a collapsing ARS spacetime the center becomes trapped at the
same comoving time at which it becomes singular, that is
\begin{equation}\label{eq:centre}
\lim_{r\to
0^+}t_h(r)=t_*=\lim_{r\to 0^+} t_s(r).
\end{equation}
\end{proposition}

In the models studied in this paper, the only singularity that can
be naked is the central ($r=0$) one. Indeed, a singularity cannot
be naked if it occurs after the formation of the apparent horizon
$t_h(r)$. But using \eqref{eq:ts-th}, we get $t_s(r)=t_h(r)$ only
if $\Psi(r,R_h(r))=0$, which happens if and only if $r=0$. Thus,
the shell labeled $r>0$ becomes trapped before becoming singular,
and hence all non--central singularities are censored. An
important consequence of this fact is that, although the
area-radius coordinates map the singularity curve to the axis
$R=0$, it is actually only the `point' $R=r=0$ that has to be
analyzed in order to establish the causal structure of the
spacetime.

Next section will be devoted to the study of the nature of the
central singularity. A singularity is either {\sl locally naked},
if it is visible to nearby observers, {\it globally naked}, if it
is visible also to far-away observers, or {\it censored}, if it is
invisible to any observer. We shall not be concerned here with the
issue of global nakedness, so that our test of cosmic censorship
will be performed on the {\it strong} cosmic censorship
hypothesis: each singularity is invisible to any observer. We call
a blackhole a singularity of this kind, although one could
conceive situations in which the singularities are locally naked
but hidden by a global event horizon, an obvious example of this
being of course the Kerr spacetime with mass to angular momentum
per unit mass ratio greater than one. However, in all the examples
so far discovered of {\it dynamical} formation of naked
singularities, one can easily attach smoothly to the region of
spacetime which contains the naked singularity a regular
asymptotic region containing no event horizon. It is, therefore,
likely that the unique version of the cosmic censorship conjecture
that can be the object of a mathematical proof is the strong one.

\section{The spectrum of endstates}\label{sec:spectrum}
\subsection{The radial null-geodesic equation}

The equation of radial null geodesics in the coordinate system
$(r,R)$ is given by
\begin{equation}\label{eq:geo-short}
\frac{\text dR}{\text dr}= u\sqrt\Delta[Y-u].
\end{equation}
Indeed, all null curves in two dimensions can be reparameterized
to become geo\-desics. Therefore, \eqref{eq:geo-short} comes from
\eqref{eq:ds} setting $\text ds^2=0$, $(\text d\theta^2 +
\sin^2\theta\,\text d\varphi^2)=0$, and
requiring the future--pointing character of the curve.

The center $R=r=0$ is (locally) naked if there exists a future
pointing local solution $R_g(r)$ of the geodesic equation which
extends back to the singularity (i.e. $R(0)=0$) and "escapes from
the apparent horizon", that is $R_g(r)>R_h(r)$ for $r>0$. We will
study in full details only the existence of {\it radial} null
geodesics emanating from the singularity. We are, however, going
to prove that if a singularity is radially censored (that is, no
radial null geodesics escape), then it is censored (see subsection
\ref{subsec:nonrad} below).

In what follows we shall need to consider {\it sub} and {\it
super} solutions of the equation \eqref{eq:geo-short}. We recall
that a function $y_0(r)$ is said to be a subsolution
(respectively supersolution)
of an ordinary differential equation of the kind
$y'=f(r,y)$ if it satisfies $y_0'\leq f(r,y_0)$ (respectively
$\geq$).

\subsection{The main theorem}

It is easy to check that in a collapsing ARS spacetime (see
Definition \ref{def:2}) the Taylor expansion of $\sqrt\Delta(r,0)$
at the center is given by
\begin{equation}\label{eq:m}
\sqrt\Delta(r,0)=\xi\, r^{n-1}+\ldots,
\end{equation}
where $\xi$ is a positive number and $n$ is a positive integer.
The following theorem shows that the causal nature of such
spacetimes is fully governed by these two quantities.

\begin{theorem}\label{teo:teo}

In a collapsing ARS spacetime, the singularity forming at the
center is locally naked if $n=1$, if $n=2$, or if $n=3$ and
$\frac\xi\alpha> \xi_{\text{c}}$ where
\begin{equation}\label{xic}
\xi_c=\frac{26+15\sqrt{3}}{2}.
\end{equation}
Otherwise the singularity is covered.
\end{theorem}

For sake of clarity, we divide the proof of the main theorem
into two parts.

\subsection{Sufficient conditions for existence}\label{sec:suff}

\begin{theorem}\label{teo:suff}
In a collapsing ARS spacetime, the singularity forming at the
center is locally naked if $n<3$ or if $n=3$ and $\frac\xi\alpha>
\xi_{\text{c}}$.
\end{theorem}
To prove the theorem we first need the following crucial result:

\begin{lemma}\label{lem:hor}
In a collapsing ARS spacetime the apparent horizon is a
supersolution of equation \eqref{eq:geo-short}.
\end{lemma}
\begin{proof}

This result can actually be proved for much more general
spacetimes (essentially, only the weak energy condition is needed)
\cite{GGM}. However, we give here a very simple proof for the
model at hand. As we have seen, we always suppose that $\alpha$ is
positive (see Definition \ref{def:2}). This implies
$2\Psi(r,0)\cong\alpha r^3$, and from the definition of the
apparent horizon $R_h(r)$, it is
\begin{equation}\label{eq:rh}
R_h(r)= 2\Psi(r,R_h(r))=2\Psi(r,0)+R_h(r) \Psi_{,R}(r,s_r)\cong
\alpha r^3,
\end{equation}
($s_r\in (0,R_h(r))$) since $\Psi_{,R}(r,\xi_r)$ is infinitesimal.
This implies $R_h'(r)\cong 3\alpha r^{2}>0$ in an open right
neighborhood of 0, whereas it is easily seen that
$u(r,R_h(r))=Y(r,R_h(r))$, implying that the right hand side of
\eqref{eq:geo-short} gives zero when evaluated at $(r,R_h)$. Hence
$R_h(r)$ is a supersolution of \eqref{eq:geo-short}.
\end{proof}
\begin{lemma}\label{lem:subsol}
The singularity forming at the center is naked if there exists
a subsolution of equation
\eqref{eq:geo-short} of the form $\tilde R(r)=x r^3$, where $x>\alpha$.
\end{lemma}
\begin{proof}
The singularity is naked if there exists a geodesic $R_g(r)$ such
that $R_g(r)>R_h(r)$ in an open right neighborhood of 0. If
$\tilde R(r)=x r^3$ is a subsolution of \eqref{eq:geo-short} and
$x>\alpha$, then $\tilde R(r)>R_h(r)$. Let $r_0>0$ and $R_g(r)$ a
geodesic through $R_g(r_0)\in]R_h(r_0),\tilde R(r_0)[$. This curve
cannot cross the subsolution from below, neither it can cross the
supersolution from above. Thus $R_g(r)$ is defined in $]0,r_0]$
and $\lim_{r\to 0^+}R_g(r)=0$. Therefore
$R_g$ is the sought geodesic.
\end{proof}

\begin{proof}[Theorem \ref{teo:suff}]
We will derive sufficient
conditions for $\tilde R(r)=xr^3$ to be a subsolution of
\eqref{eq:geo-short} with $x>\alpha$; the result will then follow
from Lemma \ref{lem:subsol}. We have
\begin{equation}\label{eq:H1}
u(r,\tilde R(r))\cong\sqrt{\frac{\alpha}{x}}
\end{equation}
(see \eqref{eq:Hh}), and
\begin{equation}\label{eq:1-H}
Y(r,\tilde R(r))-u(r,\tilde R(r))\cong 1 -\sqrt{\frac\alpha x},
\end{equation}
since $Y(0,0)=1$.

Now, from \eqref{eq:Delta-expr},
\begin{equation}\label{eq:Delta-rho}
\sqrt\Delta(r,\tilde R(r))=\sqrt\Delta(r,0)-
\int_0^{xr^3}\frac{1}{Y(r,\sigma)}\diff\Hh
r(r,\sigma)\,\text d\sigma.
\end{equation}
But
\begin{multline*}
\int_0^{xr^3}\frac{1}{Y(r,\sigma)}\diff\Hh
r(r,\sigma)\,\text d\sigma=
-\int_0^{xr^3}
\frac{\sqrt\sigma\left[\sum_{i+j=3}i\,h_{ij}r^{i-1}\sigma^j+\ldots\right]}
{2{Y(r,\sigma)}\left[\sum_{i+j=3}h_{ij}r^i\sigma^j+\ldots\right]^{3/2}}
\,\text d\sigma\cong\\
\cong-\int_0^x\frac{r^{3/2}\sqrt\tau\left[\sum_{i+j=3}i\,h_{ij}r^{i-1+3j}
\tau^j
\right]}{2\left[\sum_{i+j=3}h_{ij}r^{i+3j}\tau^j\right]^{3/2}}\,
r^3\text d\tau\cong
%-\int_0^x\frac{3 r^{3/2}r^5\,\alpha}{2\alpha^{3/2}r^{9/2}}\,\sqrt\tau\text
%d\tau=
-\frac{r^{2}}{\sqrt\alpha}\tau^{3/2}\vert_{\tau=0}^{\tau=x}=
-\sqrt{\frac x\alpha} x r^{2}
\end{multline*}
where the variable change $\sigma=r^3\tau$ has been performed, and
we recall that $\alpha=h_{30}$.
Then
\begin{equation}\label{eq:Delta-rho1}
\sqrt\Delta(r,\tilde R(r))\cong \xi\,r^{n-1}+\sqrt{\frac x\alpha}x r^{2}.
\end{equation}
The curve $\tilde R$ is certainly a subsolution of \eqref{eq:geo-short}
if the following inequality holds:
\begin{equation}\label{eq:root1}
3\,x\, r^{2}< \left(1-\sqrt{\frac\alpha
x}\right)\left(\sqrt{\frac\alpha x}\xi\, r^{n-1}+ x r^{2}\right).
\end{equation}
This inequality  holds always if $n=1$ or if $n=2$; namely, the
term on the right hand side is a positive function that behaves like
$r^m$
whenever $x>\alpha$. If $n=3$
condition \eqref{eq:root1} is equivalent to
\begin{equation}\label{eq:root2}
S(x,\frac\xi 3)< 0,
\end{equation}
where
\begin{equation}\label{eq:rootdef}
S(x,p)\equiv 2 x^2 + \sqrt\alpha x^{3/2} - 3 p\,\sqrt\alpha\, x^{1/2}+
3 p\,\alpha ,
\end{equation}
and using standard techniques it
can be seen that \eqref{eq:root2} holds for some $x>\alpha$ if and only
if
\begin{equation}\label{eq:acrit}
\frac\xi\alpha> \frac{26+15\sqrt 3}{2}=\xi_c \ .
\end{equation}
\end{proof}

\subsection{Necessary conditions for existence}\label{sec:nec}
\begin{theorem}\label{teo:nec}
In an collapsing ARS spacetime, if the singularity forming at the
center is locally naked then $n<3$ or, $n=3$ and
$\frac\xi\alpha\ge\xi_c$.
\end{theorem}

To show the theorem, essentially adapting an argument exploited in
\cite{GM} for dust solutions, we need the following Lemma.

\begin{lemma}\label{lem:bound}
In a collapsing ARS spacetime, if a curve $\tilde R(r)$ is a
geodesic emanating from the central singularity such that
\[
t_{\tilde R}(r):=\int_{\tilde R(r)}^r e^{-\nu}\Hh(r,\sigma)\,\text d\sigma
\]
verifies $\lim_{r\to 0^+}t_{\tilde R}(r)=t_0=\lim_{r\to 0^+}t_s(r)$,
then
\begin{equation}\label{eq:bound}
\lim_{r\to 0^+}\frac{\tilde R(r)}{r}=0.
\end{equation}
\end{lemma}
\begin{proof}
We have $\lim_{r\to 0^+}(t_s(r)-t_{\tilde R}(r))=0$. Recalling that
\[
t_s(r)=\int_0^r e^{-\nu}\Hh(r,\sigma)\,\text d\sigma,
\]
we also have
\begin{multline*}
t_s(r)-t_{\tilde R}(r)=\int_0^{\tilde R(r)}e^{-\nu}\Hh(r,\sigma)\,\text
d\sigma
%=\int_0^{\tilde R(r)/{r}}e^{-\nu}\Hh(r,r\tau)r\,\text d\tau\cong\\
\cong
\int_0^{\tilde R(r)/{r}}\frac{e^{-\nu}\,\sqrt\tau}
{\left(\sum_{i+j=3}h_{ij}\tau^j\right)^{1/2}}\,\text
d\tau,
\end{multline*}
and since $\nu$ is bounded this quantity must be infinitesimal, so
that  \eqref{eq:bound} must hold.
\end{proof}

\begin{proof}[Theorem \ref{teo:nec}]
Let $\tilde R(r)=x(r) r^3$ be a geodesic such that $x(r)>\alpha$, and
$\tilde R(0)=0$. Using Lemma \ref{lem:bound}, $x(r) r^2$ must be
infinitesimal, so
it is a straightforward calculation to verify that
\begin{align}
&u(r,\tilde R(r))\cong\sqrt{\frac{\alpha}{x(r)}},\label{eq:H2}\\
&Y(r,\tilde R(r))-u(r,\tilde R(r))\cong 1-\sqrt{\frac{\alpha}{x(r)}},\label{eq:1-H2}
\end{align}
whereas, using the same arguments as in \eqref{eq:Delta-rho1}, it is
\begin{multline}\label{eq:Delta-rho2}
\sqrt\Delta(r,\tilde R(r))=\sqrt\Delta(r,0)-
\int_0^{x(r)r^3}\frac{1}{Y(r,\sigma)}\diff\Hh
r(r,\sigma)\,\text d\sigma\cong\\
\cong \xi\,r^{n-1}+\sqrt{\frac{x(r)}\alpha}x(r) r^{2}.
\end{multline}
Since $\tilde R(r)$ is a geodesic, \eqref{eq:geo-short}  yields
\begin{equation}\label{eq:geo3}
x'(r)r\cong\left(1-\sqrt{\frac\alpha{x(r)}}\right)
\left[\sqrt{\frac\alpha{x(r)}}\xi r^{n-3}+ x(r)\right]-3\,x(r).
\end{equation}
By contradiction, let us first assume $n>3$. Therefore, if $x(r)$
went to $+\infty$ for $r\to 0^+$, using \eqref{eq:geo3} it would
be
\[
x'(r)=-\frac {2}{r} x(r)\psi(r)+\varphi(r),
\]
where $\varphi(r)$ is a bounded function and $\psi(0)=1$. Using
comparison theorems for ODE's there would exist $\lambda>0$ such that
$x(r)\ge \frac \lambda{r^{2}}$ and so $\tilde R(r)\ge\lambda r$, which
is in contradiction with Lemma \ref{lem:bound}. If $\lim_{r\to
0^+}x(r)=l$, with $l\in(0,+\infty)$, then a straightforward
calculation of the limit of both sides in \eqref{eq:geo3} yields a
contradiction as well, because $x'(r)\cong -\frac{k\,x(r)}r$ for
some positive constant $k$. If $x(r)$ does not have limit, there
exists a sequence $r_m$ of local minima for $x(r)$ such that
$r_m\to 0$ for $m\to\infty$, $x'(r_m)=0$ and $x(r_m)$ is bounded.
Evaluating \eqref{eq:geo3} for $r=r_m$, and taking the limit of
both sides for $m\to\infty$ we get a contradiction once again.
Then $n\le 3$. In the case $n=3$, equation \eqref{eq:geo3} can be
written as
\begin{equation}\label{eq:geo4}
x'(r)
r=-\frac{S(x(r),\frac\xi 3)}{x(r)}+\ldots,
\end{equation}
where $S(x,p)$ was defined in \eqref{eq:rootdef}. Arguing as
before, $x(r)$ cannot go to $+\infty$ as $r$ goes to 0 since
$\tilde R(r)$ would be bounded from below by $\lambda r$ for some
$\lambda>0$. If $\lim_{r\to 0^+} x(r)=l<+\infty$ then
$S(x(r),\frac\xi 3)\to S(l,\frac\xi 3)$ and, from \eqref{eq:geo4},
$S(l,\frac\xi 3)=0$. Since $l\ge\alpha$, equation \eqref{eq:acrit}
implies $\frac\xi\alpha\ge\xi_c$. If $x(r)$ does not have limit,
as before there exists a sequence $r_m\to 0$ of local minima for
$x(r)$ such that $x'(r_m)=0$ and $x(r_m)$ is bounded, and this
yields, using \eqref{eq:geo4}, that $S(x(r_m),\frac\xi 3)\to 0$,
that is $x(r_m)$ converges to a root of $S(x,\frac\xi 3)=0$
greater that $\alpha$, which is possible again only if
$\frac\xi\alpha\ge\xi_c$.
\end{proof}

\subsection{Non-radial geodesics}\label{subsec:nonrad}

We have limited our analysis to radial null geo\-de\-sics.
However, we will now show that, if no radial null geodesic escapes
from the singularity, then no null geodesic escapes at all. In
other words, we show that a radially censored singularity is
censored (in the case of dust spacetimes, this result was first
given by Nolan and Mena \cite{Nolan}).

Let by contradiction ${\tilde R}(r)$ be a non radial null geodesics
escaping from the center, that we will suppose radially censored.
Arguing in a similar way as for recovering \eqref{eq:geo-short},
we have that ${\tilde R}(r)$ solves equation
\begin{equation}\label{eq:nonrad}
\frac{\text d{\tilde R}}{\text
dr}=-u^2\left[B+\sqrt{\Delta+CL^2/R^2}\right],
\end{equation}
where $L^2$  is the conserved angular momentum. Since $C$ is
positive we have
\[
\frac{\text d{\tilde R}}{\text dr}\le
-u^2\left[B+\sqrt\Delta\right],
\]
that is ${\tilde R}(r)$ is a subsolution of the null radial geodesic
equation \eqref{eq:geo-short}. By hypothesis ${\tilde R}(r)>R_h(r)$
for $r>0$, and ${\tilde R}(0)=R_h(0)$. Then a comparison  argument
in ODE similar to the one exploited in Lemma \ref{lem:subsol}
ensures the existence of a radial null geodesic, which is a
contradiction. Thus, the following holds true:
\begin{proposition}\label{prop:nonrad}
Any singularity radially censored is censored.
\end{proposition}
We emphasize that this is a general result, not depending on the
class of solutions we are dealing with, but only on the spherical
symmetry of the model.

\subsection{Physical interpretation of the results}\label{subsec:ex}

As we have seen, although the mathematical structure of the
solutions of the Einstein field equations and the way in which
this structure governs the properties of the differential equation
of radial geodesics is extremely intricate, our final results are
nevertheless extremely simple: what governs the whole machinery is
just the first term of the Taylor expansion of $\sqrt\Delta (r,0)$
near the center (Theorem \ref{teo:teo}). To understand what this
results says in physical term, it is convenient to write
$\sqrt\Delta(r,0)$ as
\begin{equation}\label{eq:Delta02}
\sqrt\Delta(r,0)=I_1(r) +I_2(r)
\end{equation}
where (see also \eqref{eq:Delta-expr})
\begin{equation}\label{eq:Delta021}
I_1(r):=\frac 1{Y(r,r)}
\left[\diff{}r\int_0^r\Hh(r,\sigma)\,\text d\sigma\right],\quad
I_2(r):= \int_0^r \gamma(r,\sigma) \diff\Hh r(r,\sigma)\,\text
d\sigma,
\end{equation}
and we have defined
$$
\gamma(r,\sigma ):=\frac{1}{Y(r,\sigma)}-\frac{1}{Y(r,r)}.
$$
The function $I_1(r)$ has the following behavior:
\begin{multline}\label{eq:Delta0}
I_1(r)=
\frac 1{Y(r,r)}\diff{}r\left(\int_0^1\Hh(r,r\tau)r\,\text
d\tau\right)
\cong\\
\cong\int_0^1 \frac{\sqrt\tau(p\,
r^{p-1}Q(\tau)+\ldots}{2\left[P(\tau)-r^p
Q(\tau)+\ldots\right]^{3/2}} \cong p\,a\,r^{p-1}+\ldots,
\end{multline}
where the following Taylor expansion through $(0,0)$ has been introduced:
\begin{equation}\label{eq:H(r,R)}
2\Psi(r,R)+R(Y^2(r,R)-1)=\sum_{i+j=3}h_{ij}r^j R^j+
\sum_{i+j=3+p}h_{ij}r^j R^j+\ldots,
\end{equation}
and
\begin{equation}\label{eq:Aba}
P(\tau)=\sum_{i+j=3}h_{ij}\tau^j,\quad
Q(\tau)=-\sum_{i+j=3+p}h_{ij}\tau^j,\quad
a=\int_0^1\frac{Q(\tau)\sqrt\tau}{2 P(\tau)^{3/2}}\,\text
d\tau.
\end{equation}
The remaining summand $I_2(r)$ is zero if $Y$ depends only on $r$,
i.e. if the acceleration $A(r,R)=Y_{,R}(r,R)$ (formula
\eqref{eq:acc}) vanishes. Its behavior is as follows:
\begin{multline*}
I_2(r)=-\int_0^r \gamma(r,\sigma )
\frac{\sqrt\sigma\left[\sum_{i+j=3}i\,h_{ij}r^{i-1}\sigma^j+\ldots\right]}
{2\left[\sum_{i+j=3}h_{ij}r^i\sigma^j+\ldots\right]^{3/2}}
\,\text d\sigma=\\
=-\int_0^1 \gamma(r,r\tau)
\frac{\sqrt\tau\left[\sum_{i+j=3}i\,h_{ij}\tau^j+\ldots\right]}
{2r\left[P(\tau)+\ldots\right]^{3/2}} \,\text d\tau.
\end{multline*}
Now, using Taylor expansion of $Y$ through the center
\[
Y(r,R)=\varphi(r)+\sum_{\substack{i+j=q+1 \\ j>0}} k_{ij} r^i
R^j+\ldots,
\]
($\varphi(r)$ contains the terms where $R$ does not appear, and
$\varphi(0)=1$), it is
$$
\gamma(r,r\tau)\cong
Y(r,r)-Y(r,r\tau)=r^{1+q}\sum_{\substack{i+j=q+1 \\ j>0}}
k_{ij}(1-\tau^j)+\ldots=S(\tau)\,r^{1+q}+\ldots,
$$
where $q$ is easily seen to be the order of the first non
vanishing term of the expansion of the acceleration near the
center. Thus, we obtain:
\[
I_2(r)=-b\,r^q+\ldots,\quad\text{where\ }
b:=\int_0^1 S(\tau)
\frac{\sqrt\tau\left[\sum_{i+j=3}i\,h_{ij}\tau^j\right]}
{2P(\tau)^{3/2}} \,\text d\tau.
\]
It follows that
\begin{equation}\label{eq:Delta022}
\sqrt\Delta(r,0)= p\,a\,r^{p-1}-b\,r^q+\ldots
\end{equation}
The index $n$ defined in formula \eqref{eq:m} is thus the
smaller\footnote{One can conceive very special cases in which the
terms exactly balance each other and the index $n$ has to be
defined at the next order.} between $p$ and $q+1$. The value of
$p$ is clearly related to the degree of inhomogeneity of the
system, since one can generate a low value of it taking a low
order of non-vanishing derivatives of the initial density profile
at the center (formula \eqref{eq:eps0}). This effect can be
related to the shear as well (see e.g. \cite{bjm,jdad}). The value
of $q$, and thus the second term, is related strictly to the
acceleration: it vanishes if the acceleration vanishes, and in any
case it does not contributes to the nature of the final state if
the system ''does not accelerate enough'' near the center. The
effect of this term on naked singularities formation can be
considered as a new feature of our models. It can, in fact, be
shown that (virtually) {\it all} the particular cases already
known in literature of formation of naked singularities in the
gravitational collapse of continua (for instance the dust and the
tangential stress model) can be retrieved as particular cases of
our main theorem, and that in all such cases the acceleration term
is negligible and the effect does not occur (of course, these
cases to not exhaust the content of Theorem \ref{teo:teo}).

\section{Discussion and conclusions}

From the very beginning, the problem of Cosmic Censorship was
shown to be linked to specific mathematical (as opposed to
physically transparent)  properties of the available arbitrary
functions, like e.g. `first derivative of initial density equal to
zero, second non-zero at the center'. In all the cases which have
been discovered so far the situation became more and more
intricate.  In the present paper, we have constructed a new class
of solutions which contains as subcases all the solutions for
which censorship has already been investigated in full details.
The new solutions add to the (rather scarce) set of spherically
symmetric spacetimes whose kinematical properties are generic. The
analysis of the structure of their singularities allowed us to
show that a general and simple pattern actually exists. This
pattern follows from Theorem \ref{teo:teo}: given any set of
regular data, and any equation of state within the considered
class, the formation of naked singularities or blackholes depends
on a sort of selection mechanism. The mechanism works as follows:
it extrapolates the value of an integer $n$ and selects the final
state according to it. The extrapolation essentially depends on
the kinematical invariants of the motion. If the resulting $n$
equals one or two or it is greater than three, the final state is
decided and has no other dependence on the data or on the matter
properties. Dimensional quantities such as e.g. value of the
derivative of the density at the center, cosmological constant,
value of the derivative of the velocity profile at the center,
profile of the state equation for tangential stresses, and so on,
play a further role only at the transition between the two
endstates, occurring at $n=3$. This role is to combine themselves
to produce a non-dimensional quantity which acts as a critical
parameter.

The (spherically symmetric) cosmic censor seems to answer to the
court, that the formation of naked singularities or blackholes is
essentially a {\it local and kinematical} phenomenon: it neither
depends (or weakly depends) on {\it what is collapsing}, nor it
depends on the details of the data characterizing {\it how it
starts collapsing}. The formation or whatsoever of blackholes or
visible singularities depends only on the kinematical properties
of the motion near the center of symmetry of the system.

\textbf{Acknowledgement.}
The authors wish to thank the referees for their
very useful comments and suggestions.

\end{document}